\documentclass[%
 reprint,
 amsmath,amssymb,
 aps,
]{revtex4-1}

\usepackage{graphicx}
\usepackage{dcolumn}
\usepackage{bm}
\usepackage{epstopdf}
\usepackage{wasysym,letltxmacro}

\newcommand{\pii}{\mathcal{P}}  
\begin{document}
\preprint{APS/123-QED}
\title{The Quantum Trihedron}
\author{A. Bendjoudi}
\email{8ahmida8@gmail.com}
\author{N. Mebarki}
\email{nnmebarki@yahoo.fr}
\affiliation{Department of Physics, Fr\`{e}res Mentouri University, Constantine Ain El Bey 25000}

\date{\today}

\begin{abstract}
The convex hull on three points in two dimensional euclidean space of three flat edges (trihedron) was studied. The Bohr-Sommerfeld quantization of the area of space is performed. It is shown that it reproduces exactly the equidistant spacing spectrum found elsewhere.
\begin{description}  
\item[PACS numbers]
04.60.Pp, 04.60.Nc, 03.65.Sq
\end{description}
\end{abstract}

\maketitle         
   
During the last years, a great interest to loop quantum gravity has been devoted \cite{1,2} where the spacetime granularity having a discrete spectrum with a rich structure has been discovered. The main feature of this new quantum theory of gravity lies in the fact that it starts from the quantization of general relativity using canonical, covariant and geometric approaches to end up with the discretization of space.   
                                                                                                                                     
In the present paper, we present a new independent approach to the granularity of space and the computation of the area spectrum. The derivation is purely semiclassical applied to the more simplest model of a grain of space, an euclidean trihedron, within the relation between the quantum polyhedra and loop gravity addressed in Ref.~\cite{6}. The role of Hamiltonian generating classical orbits is played by the area of space.	Together with the Planck hypothesis on the discreteness of 2d symplectic areas, we reproduce the well-known semiclassical spectrum of the area of space discussed elsewhere: \emph{the equidistant spacing spectrum}.                                 

Mathematically, the quantum polyhedra are the result of the quantization on the classical phase spaces of polyhedra, which are called \emph{polyhedra space of shapes}. These spaces can be fully explored via two key results \cite{4,5}:    
\begin{enumerate} 
\item a theorem due to Minkowski which states that  for $h$ non-co-planar unit vectors $\left\{ \vec{n}_i \right\}$ and $h$ positive numbers $\left\{ A_i \right\}$ such that the closure  condition $  \sum^{h}_{k=1} A_{k} \vec{n}_{k} = \vec{0}  $ holds, then the areas $ \left\{ \vec{A}_{i} = A_{i}\vec{n}_{i}  \right\} $ fully characterize the shapes of a geometrical object called polyhedron (up to rotations and translations, this polyhedron is unique). This allows to define the space  of shapes of polyhedron $ \pii_{h}$ to be (up to $SO(3)$ rotation)                 
\begin{equation} 
\pii_{h}=\left\{\vec{A}_{k}, k=1,..., h ~|~ \sum_{k=1}^{h} A_{k} \vec{n}_{k} = \vec{0} , \left|\vec{{A}_{k}}\right|= A_{k}\right\}; 
\label{eq1}    
\end{equation}
\item the so-called Kapovich-Millson (KM) phase space: As it is shown in Refs.\cite{5,6}, the set $\pii$ has a symplectic structure. The Poisson brackets between two arbitrary functions $ F(A_1, A_2, \cdots, A_h) $ and $ G(A_1, A_2, \cdots, A_h)$ can be defined via  
\begin{equation}  
\left\{F, G\right\}=\sum_{i=1}^{h}\vec{A}_{i}\cdot\left(\sum^{3}_{a=1}\frac{\partial F}{\partial A_{ia}} \vec{u}_{ia}\times \sum^{3}_{b=1}\frac{\partial G}{\partial A_{ib}} \vec{u}_{ib}\right) 
\label{eq2}.    
\end{equation}            
\end{enumerate}
where $\vec{A}_i = \sum^{3}_{a=1}A_{ia} \vec{u}_{ia}$ such that $\left\{\vec{u}_{ia}\right\}$ are the three cartesian unit vectors and $\left\{A_{ia}\right\}$ the corresponding components (note that $\left\{\vec{A}_{i}\right\}$ are vectors in the usual 3d euclidean space representing the areas of the polyhedron, each vector has three components in the cartesian coordinates)               

The variables that satisfy the equations $ \left\{ p_{m}, q_{n}\right\} = \delta _{mn}  $ are defined as follows: the momenta variables are the norms $ \left|\vec{p}_{m}\right| = \left|\sum_{l=1}^{m+1} \vec{A}_{l}\right| $ where $ m= 1, \ldots, h-3$, the coordinates variables is the angles between the vectors $ \vec{p}_{m} \times  \vec{A}_{m+1} $ and  $ \vec{p}_{m} \times  \vec{A}_{m+2} $ (for more details see Ref.~\cite{6}).For our case, a classical tetrahedron, the momentum variable  is the norm $ p=\left| \vec{A}_{1}+\vec{A}_{2} \right| $. The coordinate variable $  \varphi $  is the angle between the two vectors  $ \vec{A}_{1} \times \vec{A}_{2}  $ and  $ \vec{A}_{3} \times \vec{A}_{4} $. The conjugated pair $(p,\varphi)$ defines the phase space of tetrahedron. Performing Bohr-Sommerfeld quantization on it results the quantum tetrahedron.  

\begin{figure}[t]          
	\centering
		\includegraphics[width=1.19in]{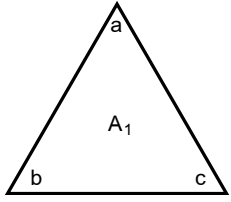}
	\caption{The convex hull on the three points $ a, b$ and $c$ in 2d euclidean space bounding a 2d area $ A_1$.}
	\label{fig1}  
\end{figure}

Now, the ground is established to proceed into the quantization. We start with investigating classical trihedra. The statement “trihedron” means the convex hull on three points (see Fig.~\ref{fig1}), which can be set in a 2d euclidean space $ \partial \xi \left( \alpha, \beta\right)$ where  $ \alpha $ and $ \beta$ are its dimensions. The corresponding edges, in fact, do not correspond to 2d areas, but rather lengths encircling a 2d area. This area is defined to be a face for a tetrahedron $ \tau$.          

\begin{figure}[ph]
\includegraphics[width=3.19in]{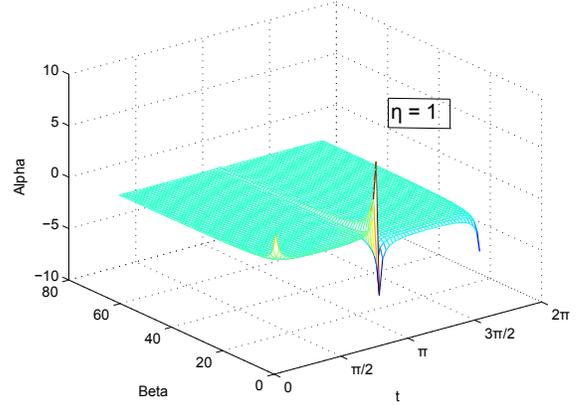}
\caption{\label{fig:epsart}The mesh of the edge $ \alpha_0$ in terms of  $\beta_0$ and $t$, it presents infinite curves between the two edges $\alpha_0$ and $\beta_0$ that satisfy $A_1=\eta = \frac{1}{2}\alpha_0 \beta_0 \sin \varphi =  1, \varphi = \left\{\varphi_0, \varphi_{\acute{0}}\right\}$}       
	\label{fig2} 
\end{figure}

Let us now consider the euclidean 3d space $ \xi \left(\alpha, \beta, \lambda\right)$ where  $ \alpha $, $ \beta$ and $ \lambda $ are its dimensions and $ \partial \xi \left(\alpha \left(\lambda\right), \beta\left(\lambda\right)\right)$ its slices, in which we localize the $ \xi$ three coordinates associated with a tetrahedron $ \tau $ by bounding them in intervals such that $0 \leq \alpha \leq \alpha_0, 0\leq \beta \leq\beta_0$ and $ 0\leq \lambda \leq \sigma $. The $ \tau $ areas can be defined by choosing a basis at a corner $c$ in $ \tau $ of three unit vectors $\vec{u}_\alpha,\vec{u}_\beta$ and $\vec{u}_\lambda$ such that $ \vec{A}_1 = \frac{1}{2} \alpha_0 \beta_0 \left(\vec{u}_\alpha \times \vec{u}_\beta\right), 
\vec{A}_2 = \frac{1}{2} \alpha_0 \sigma \left( \vec{u}_\alpha \times \vec{u}_\lambda\right), \vec{A}_3 = \frac{1}{2} \beta_0 \sigma\left(\vec{u}_\lambda \times \vec{u}_\beta\right)$ and $ \vec{A}_4 = - \left(\vec{A}_1+\vec{A}_2+\vec{A}_3\right)$. At $c$, three edges intersect and they can be defined as 
$ \vec{l}_1 = \alpha_0 \vec{u}_\alpha,  \vec{l}_2 = \beta_0 \vec{u}_\beta$ and  $\vec{l}_3 = \sigma \vec{u}_\lambda$.                       
The task now is to find the boundaries of the KM phase space, that is the phase space of the $ \tau$ areas. To do so, we consider for simplicity (of course the general argument does not depend on this choice) the following: 
$ \left|\vec{A}_2\right|=\left|\vec{A}_3\right|$ and $ \vec{u}_\lambda \parallel  \vec{A}_1$. Thus for small values of $\sigma$, $ \left|\vec{A}_1\right|\cong \left|\vec{A}_4\right|$. Using this and the relation                       
\begin{equation}   
Q = \frac{8}{9} \frac{\Delta\overline{\Delta}}{p} \sin \varphi  
\label{eq3}  
\end{equation} 
where $ \Delta$ and $\overline{\Delta}$ are the areas encircled by $\vec{A}_1, \vec{A}_2$, $\vec{p}=\vec{A}_1+\vec{A}_2$ and $\vec{A}_3, \vec{A}_4$, $-\vec{p}=\vec{A}_3+\vec{A}_4$ respectively and $Q$ is the squared volume of $\tau$, one can show that the KM phase space of $\tau$  reduces in the boundary $ \partial \xi \left( \alpha \left(\lambda\right), \beta\left(\lambda\right)\right)$ (for the area $\left|\vec{A}_1\right|$ ) to the points $P_1$ and $P_2$ given by      
\begin{eqnarray}
 P_1 &=& \left(\eta, \varphi_0 = \arcsin \frac{2\eta}{\alpha_0\beta_0}\right), \\ 
P_2 &=& \left(\eta, \varphi_{\acute{0}} = \arcsin \frac{-2\eta}{\alpha_0\beta_0}\right)  
\label{eq4} 
\end{eqnarray}
where $ \eta = \left|\vec{A}_1\right|$ such that $\varphi_0, \varphi_{\acute{0}} \neq n \pi$, $n = 0,1,2 \cdots$. It is to be noted that this symplectic space is the KM phase space of $\tau$ corresponding to the limit $\sigma\longmapsto 0$. Considering the relation between $\sigma$ and the areas of $\tau$ one can visualize the structure of $\tau$ at the boundary. Note that $P_1$ and $P_2$ are functions of the area-configuration and at each point on classical orbits the area corresponds to a given configuration. The variables that define the configuration are the lengths of $\alpha_0$ and $\beta_0$.  In this space, the area $\eta$ plays the role of the Hamiltonian generating classical orbits and it is, in addition, constant. This is nicely consistent with the solutions of the equations of dynamics, which   are                                                 
\begin{eqnarray}
 \frac{d\eta}{dt} = \left\{\eta,\eta\right\} &=& 0, \\   
 \frac{d\varphi}{dt} = \left\{\varphi,\eta\right\} &=& 1     
\label{eq5}
\end{eqnarray}
where $t$ is defined to be the canonically conjugate to the Hamiltonian, i.e. $\varphi=t$. In addition, one can use this equations to study the evolution of the lengths $\alpha_0$ and $\beta_0$ over $t$ and get the solution                     
\begin{equation}   
 \alpha_0\left(t\right) \beta_0 \left(t\right) = \frac{\omega}{\sin t}       
\label{eq7}    
\end{equation}     
where $\omega$ is a constant of integration, i.e. constant along classical orbits, equal to twice the area bounded by $\alpha_0 \left(t\right)$ and $\beta_0\left(t\right)$, and $0\leq t \leq 2\pi$ such that $t \neq n \pi$ (for integer $n$). This analysis shows how the lengths of the triangle edges, its area and the angle between the edges vary consistently in a quantum fashion. Fig.~\ref{fig2} represents the space of configuration of the area $\eta$ in terms of its edges lengths.                                          

Let us now proceed into the quantization, the basic idea in doing so is the Planck hypothesis: \emph{symplectic areas in the phase space of a system vary in discrete steps; they are $2\pi\left(n + \frac{1}{2}\right)$ times the Plank constant}. Bohr-Sommerfeld orbits reduce to closed circle-orbits on which the points $P_1$ and $P_2$ are defined. Bohr-Sommerfeld quantization can be performed as                                                       
\begin{equation}     
 \oint \eta d\varphi = 2\pi h \left( n +\frac{1}{2}\right) = 2\pi\eta          
\label{eq8}   
\end{equation}
where the integer $n$ stands for the levels of the Hamiltonian (levels of $\eta$). In performing the integration in Eq. (9), we have proceeded as follows: (a) the area $\eta$ plays the role of the Hamiltonian generating classical orbits, therefore it is constant along closed orbits thus (b) the integration reduces to computing the total change in $\varphi$ in the motion along the closed orbits. In computing the total change, $\varphi$ varies on the points on the orbits except for the singular points $\varphi=n \pi$, because they correspond to meaningless cases; the area, for these points, becomes 1d line. Thus, they (the points: $\varphi = 0~\rm{and}~\pi$) are not relevant with the present quantization procedure. It is worth to mention that the two points $\varphi=0$ and $\varphi=2\pi$ represent a periodicity, i.e. they are set at the same point on the orbit under consideration. 

The last equation tells us that choosing a value for $\left|\vec{A}_1\right|$ in the quantum tetrahedron amounts to pick up a value for $n$. This level corresponds to the $su(2)$ algebra value, that is $n=j$. This reproduces the well-known equidistant spacing spectrum for the area of space found and discussed semiclassically elsewhere    
 \begin{equation}   
 \eta = h\left(j+\frac{1}{2}\right)       
\label{eq9}.   
\end{equation}

In the context of KM phase space, this formula was assumed in Ref.~\cite{8}. In Ref.~\cite{3}, the same formula was derived by lifting the problem of angular momenta addition into the space of Schwinger’s oscillators.  Furthermore,  it was shown in Ref.~\cite{9} that assigning the spectrum of this area to the links variables in loop gravity can reproduce both the thermodynamics and the quasinormal mode properties of black holes. However, this formula has a long history starting  from the Ponzano and Regge paper of Ref.~\cite{7} to the present work \cite{3}. In comparing this with the quantization of the area proved canonically in loop gravity \cite{10,11} one can see  that the quantitative agreement is extremely well in the large values of areas.

It is to be noted that Bohr-Sommerfeld quantization rule has a domain of validity and can be seen, in fact, as the semiclassical approximation to canonical quantum mechanics. After the birth of quantum mechanics by the Planck hypothesis, this rule has become the basic principle of what is known as \emph{old quantum theory}. However, this rule has well-known limits: (a) it provides no means to calculate the intensities of spectral lines, (b) it fails to explain the \emph{anomalous Zeeman effect} where the spin of electrons is taken into account and (c) it cannot quantize \emph{chaotic systems} where the trajectories generated by the Hamiltonian of the system under study are neither closed nor periodic. Yet this rule still important, especially in studying the semiclassical approximation of quantum states in loop quantum cosmology. Note that neither spectral lines intensities nor anomalous Zeeman effect can be studied practically in handling the quantum spacetime structure due to the fact that quantum gravity still a theoretical notion far from being at the experience stage. The chaotic systems can be studied theoretically, which get arisen in systems of higher phase space dimension such as the quantum pentahedron or more generally quantum polyhedron.                               

In summary, we have studied semiclassically the quantum theory of trihedron, the convex hull on three points in 2d euclidean space. We have quantized semiclassically its volume (2d area). Our main result is summarized in Eq.~(10). The correspondence in this quantization with the quantum tetrahedron is that the three lengths that bound $\eta$ correspond to the four areas that surround the $\tau$ volume.

A remarkable feature characterizing the result of the present work is that it provides new insights into our understanding to the quantum polyhedra and its relevance to Bohr-Sommerfeld quantization, especially the practical dealing with loop gravity and its applications. More investigations are under study.

\end{document}